\documentclass[10pt,a4paper]{article}
\usepackage{amsmath,amssymb,latexsym,amscd,amsthm}
\usepackage[T1]{fontenc}
\usepackage[english]{babel}
\usepackage{graphicx}

\begin{document}

\title{Separability criteria based on the Bloch representation of density matrices}

\author{Julio I. de Vicente\footnote{E-mail address: jdvicent@math.uc3m.es}\\Departamento de Matem\'aticas \\ Universidad Carlos III de Madrid \\
Avda.\ de la Universidad 30, E-28911 Legan\'es, Madrid, Spain}

\date{}

\maketitle

\begin{abstract}
We study the separability of bipartite quantum systems in arbitrary
dimensions using the Bloch representation of their density matrix.
This approach enables us to find an alternative characterization of
the separability problem, from which we derive a necessary condition
and sufficient conditions for separability. For a certain class of
states the necessary condition and a sufficient condition turn out
to be equivalent, therefore yielding a necessary and sufficient
condition. The proofs of the sufficient conditions are constructive,
thus providing decompositions in pure product states for the states
that satisfy them. We provide examples that show the ability of
these conditions to detect entanglement. In particular, the
necessary condition is proved to be strong enough to detect bound
entangled states.
\end{abstract}

\section{Introduction}

Let $\rho$ denote the density operator, acting on the
finite-dimensional Hilbert space $H=H_A\otimes H_B$, which describes
the state of two quantum systems $A$ and $B$. The state is said to
be separable if $\rho$ can be written as a convex combination of
product vectors \cite{Wer}, i.e.
\begin{equation}\label{separable}
\rho = \sum_ip_i|\phi_i,\varphi_i\rangle\langle\phi_i,\varphi_i| =
\sum_i p_i \, \rho^A_i\otimes\rho^B_i \,,
\end{equation}
where $0 \leq p_i \leq 1$, $\sum_i p_i = 1$, and
$|\phi_i,\varphi_i\rangle=|\phi_i\rangle_A\otimes|\varphi_i\rangle_B$
($|\phi\rangle_A\in H_A$ and $|\varphi\rangle_B\in H_B$).

If $\rho$ cannot be written as in Eq.\ (\ref{separable}), then the
state is said to be entangled. Entanglement is responsible for many
of the striking features of quantum theory and, therefore, it has
been an object of special attention. Since the early years of
quantum mechanics, it has been present in many of the debates
regarding the foundations and implications of the theory (see e.g.
\cite{epr}), but in the last ten years this interest has greatly
increased, specially from a practical point of view, because
entanglement is an essential ingredient in the applications of
quantum information theory, such as quantum cryptography, dense
coding, teleportation and quantum computation \cite{Nie,Bou}. As a
consequence, much effort has been devoted to the so-called
separability problem, which consists in finding mathematical
conditions which provide a practical way to check whether a given
state is entangled or not, since it is in general very hard to
verify if a decomposition according to the definition of
separability (\ref{separable}) exists. Up to now, a conclusive
answer to the separability question can only be given when $\dim
H_A=2$ and $\dim H_B=2$ or $\dim H_B=3$, in which case the
Peres-Horodecki criterion \cite{Per,Hor} establishes that $\rho$ is
separable if and only if its partial transpose (i.e. transpose with
respect to one of the subsystems) is positive. For higher dimensions
this is just a necessary condition \cite{Hor}, since there exist
entangled states with positive partial transpose (PPT) which are
bound entangled (i.e. their entanglement cannot be distilled to the
singlet form). Therefore the separability problem remains open. Much
subsequent work has been devoted to finding necessary conditions for
separability (see for example
\cite{iso,NieKem,Rud1,Che,Hof,Guh,deV}), given that they can assure
the presence of entanglement in experiments and that, in principle,
they might complement the strong Peres-Horodecki criterion by
detecting PPT entanglement. Nevertheless, there also exist a great
variety of sufficient conditions (such as \cite{ZHSL,Gur}),
non-operational necessary and sufficient conditions (see for
instance \cite{Hor,Ter,Wu}), or necessary and sufficient conditions
which apply to restricted sets such as low-rank density matrices
\cite{lowrank}. Furthermore, given a generic separable density
matrix it is not known how to decompose it according to Eq.\
(\ref{separable}) save for the ($2\times2$)-dimensional case
\cite{Woo,San}. The (approximate) separability problem is NP-hard
\cite{NP}, but several authors have devised nontrivial algorithms
for it (see \cite{Ian3} for a survey).

In this paper we derive a necessary condition and three sufficient
conditions for the separability of bipartite quantum systems of
arbitrary dimensions. The proofs of the latter conditions are
constructive, so they provide decompositions in product states as in
Eq.\ (\ref{separable}) for the separable states that fulfill them.
Our results are obtained using the Bloch representation of density
matrices, which has been used in previous works to characterize the
separability of a certain class of bipartite qubit states
\cite{Hor2} and to study the separability of bipartite states near
the maximally mixed one \cite{Cav,Run}. The approach presented here
is different and more general. We will also provide examples that
show the usefulness of the conditions derived here. Remarkably, the
necessary condition is strong enough to detect PPT entangled states.
Finally, we will compare this condition to the so-called computable
cross-norm \cite{Rud1} or realignment \cite{Che} (CCNR) criterion,
which exhibits a powerful PPT entanglement detection capability,
showing that for a certain class of states our condition is
stronger.

\section{Bloch Representation}

$N$-level quantum states are described by density operators, i.e.
unit trace Hermitian positive semidefinite linear operators, which
act on the Hilbert space $H\simeq\mathbb{C}^N$. The Hermitian
operators acting on $H$ constitute themselves a Hilbert space, the
so-called Hilbert-Schmidt space $HS(H)$, with inner product
$\langle\rho,\tau\rangle_{HS}=\textrm{Tr}(\rho^\dag\tau)$.
Accordingly, the density operators can be expanded by any basis of
this space. In particular, we can choose to expand $\rho$ in terms
of the identity operator $I_N$ and the traceless Hermitian
generators of $SU(N)$ $\lambda_i$ $(i=1,2,\ldots,N^2-1)$,
\begin{equation}\label{bloch}
\rho=\frac{1}{N}(I_N+r_i\lambda_i),
\end{equation}
where, as we shall do throughout this paper, we adhere to the
convention of summation over repeated indices. The generators of
$SU(N)$ satisfy the orthogonality relation
\begin{equation}\label{ortogonalidad}
\langle\lambda_i,\lambda_j\rangle_{HS}=\textrm{Tr}(\lambda_i\lambda_j)=2\delta_{ij},
\end{equation}
and they are characterized by the structure constants of the
corresponding Lie algebra, $f_{ijk}$ and $g_{ijk}$, which are,
respectively, completely antisymmetric and completely symmetric,
\begin{equation}\label{algebrasu}
\lambda_i\lambda_j=\frac{2}{N}\delta_{ij}I_N+if_{ijk}\lambda_k+g_{ijk}\lambda_k.
\end{equation}
The generators can be easily constructed from any orthonormal basis
$\{|a\rangle\}_{a=0}^{N-1}$ in $H$ \cite{Hio}. Let $l,j,k$ be
indices such that $0\leq l\leq N-2$ and $0\leq j<k\leq N-1$. Then,
when $i=1,\ldots,N-1$
\begin{equation}\label{generadoresa}
\lambda_i=w_l\equiv\sqrt{\frac{2}{(l+1)(l+2)}}\left(\sum_{a=0}^l|a\rangle\langle
a|-(l+1)|l+1\rangle\langle l+1|\right),\quad
\end{equation}
while when $i=N,\ldots,(N+2)(N-1)/2$
\begin{equation}\label{generadoresb}
\lambda_i=u_{jk}\equiv|j\rangle\langle k|+|k\rangle\langle j|,
\end{equation}
and when $i=N(N+1)/2,\ldots,N^2-1$
\begin{equation}\label{generadoresc}
\lambda_i=v_{jk}\equiv-i(|j\rangle\langle k|-|k\rangle\langle j|).
\end{equation}

The orthogonality relation (\ref{ortogonalidad}) implies that the
coefficients in (\ref{bloch}) are given by
\begin{equation}\label{ri}
r_i=\frac{N}{2}\textrm{Tr}(\rho\lambda_i).
\end{equation}
Notice that the coefficient of $I_N$ is fixed due to the unit trace
condition. The vector $\textbf{r}=(r_1 r_2\cdots r_{N^2-1})^t\in
\mathbb{R}^{N^2-1}$, which completely characterizes the density
operator, is called Bloch vector or coherence vector. The
representation (\ref{bloch}) was introduced by Bloch \cite{Blo} in
the $N=2$ case and generalized to arbitrary dimensions in
\cite{Hio}. It has an interesting appeal from the experimentalist
point of view, since in this way it becomes clear how the density
operator can be constructed from the expectation values of the
operators $\lambda_i$,
\begin{equation}
\langle\lambda_i\rangle=\textrm{Tr}(\rho\lambda_i)=\frac{2}{N}r_i.
\end{equation}

As we have seen, every density operator admits a representation as
in Eq.\ (\ref{bloch}); however, the converse is not true. A matrix
of the form (\ref{bloch}) is of unit trace and Hermitian, but it
might not be positive semidefinite, so to guarantee this property
further restrictions must be added to the coherence vector. The set
of all the Bloch vectors that constitute a density operator is known
as the Bloch-vector space $B(\mathbb{R}^{N^2-1})$. It is widely
known that in the case $N=2$ this space equals the unit ball in
$\mathbb{R}^3$ and pure states are represented by vectors on the
unit sphere. The problem of determining $B(\mathbb{R}^{N^2-1})$ when
$N\geq3$ is still open and a subject of current research (see for
example \cite{KimKos} and references therein). However, many of its
properties are known. For instance, using Eq.\ (\ref{algebrasu}),
one finds that for pure states ($\rho^2=\rho$) it must hold
\begin{equation}\label{rpuro}
||\textbf{r}||_2=\sqrt{\frac{N(N-1)}{2}},\quad
r_ir_jg_{ijk}=(N-2)r_k,
\end{equation}
where $||\cdot||_2$ is the Euclidean norm on $\mathbb{R}^{N^2-1}$.

In the case of mixed states, the conditions that the coherence
vector must satisfy in order to represent a density operator have
been recently provided in \cite{Kim,Byr}. Regrettably, their
mathematical expression is rather cumbersome. It is also known
\cite{Har,Kos} that $B(\mathbb{R}^{N^2-1})$ is a subset of the ball
$D_R(\mathbb{R}^{N^2-1})$ of radius $R=\sqrt{\frac{N(N-1)}{2}}$,
which is the minimum ball containing it, and that the ball
$D_r(\mathbb{R}^{N^2-1})$ of radius $r=\sqrt{\frac{N}{2(N-1)}}$ is
included in $B(\mathbb{R}^{N^2-1})$. That is,
\begin{equation}\label{inclusion}
D_r(\mathbb{R}^{N^2-1})\subseteq B(\mathbb{R}^{N^2-1})\subseteq
D_R(\mathbb{R}^{N^2-1}).
\end{equation}

In the case of bipartite quantum systems of dimensions $M\times N$
($H\simeq\mathbb{C}^M\otimes\mathbb{C}^N$) composed of subsystems
$A$ and $B$, we can analogously represent the density operators
as\footnote{This representation is sometimes referred in the
literature as Fano form (see e.\ g.\ \cite{BZ}), since this author
was the first to consider it \cite{Fan}.}
\begin{equation}\label{bipartitebloch}
\rho=\frac{1}{MN}(I_M\otimes I_N+r_i\lambda_i\otimes
I_N+s_jI_M\otimes\tilde{\lambda}_j+t_{ij}\lambda_i\otimes\tilde{\lambda}_j),
\end{equation}
where $\lambda_i$ ($\tilde{\lambda}_j$) are the generators of
$SU(M)$ ($SU(N)$). Notice that $\textbf{r}\in \mathbb{R}^{M^2-1}$
and $\textbf{s}\in \mathbb{R}^{N^2-1}$ are the coherence vectors of
the subsystems, so that they can be determined locally,
\begin{equation}
\rho_A=\textrm{Tr}_B\rho=\frac{1}{M}(I_M+r_i\lambda_i),\quad\rho_B=\textrm{Tr}_A\rho=\frac{1}{N}(I_N+s_i\tilde{\lambda}_i).
\end{equation}
The coefficients $t_{ij}$, responsible for the possible
correlations, form the real matrix $T\in \mathbb{R}^{(M^2-1)\times
(N^2-1)}$, and, as before, they can be easily obtained by
\begin{equation}\label{T}
t_{ij}=\frac{MN}{4}\textrm{Tr}(\rho\lambda_i\otimes\tilde{\lambda}_j)=\frac{MN}{4}\langle\lambda_i\otimes\tilde{\lambda}_j\rangle.
\end{equation}

\section{Separability Conditions from the Bloch Representation}

The Bloch representation of bipartite quantum systems
(\ref{bipartitebloch}) allows us to find a simple characterization
of separability for pure states.

\vspace*{12pt} \noindent {\bf Proposition~1:} A pure bipartite
quantum state with Bloch representation (\ref{bipartitebloch}) is
separable if and only if
\begin{equation}\label{pureseparable}
T=\textbf{r\,s}^t
\end{equation}
holds.

\vspace*{12pt} \noindent {\bf Proof:} Simply notice that Eq.\
(\ref{bipartitebloch}) can be rewritten as
\begin{equation}
\rho=\rho_A\otimes\rho_B+\frac{1}{MN}[(t_{ij}-r_is_j)\lambda_i\otimes\tilde{\lambda}_j].
\end{equation}
Since the $\lambda_i\otimes\tilde{\lambda}_j$ are linearly
independent, $(t_{ij}-r_is_j)\lambda_i\otimes\tilde{\lambda}_j=0$ if
and only if $t_{ij}-r_is_j=0$ $\forall\, i,j$.\hfill$\square$

\vspace*{12pt} \noindent {\bf Remark~1:} In the case of mixed
states, Eq.\ (\ref{pureseparable}) provides a sufficient condition
for separability, since then $\rho=\rho_A\otimes\rho_B$.
\vspace*{12pt}

Attending to Proposition 1, we can characterize separability from
the Bloch representation point of view in the following terms:

\emph{A bipartite quantum state with Bloch representation
(\ref{bipartitebloch}) is separable if and only if there exist
vectors} $\textbf{u}_i\in \mathbb{R}^{M^2-1}$ \emph{and}
$\textbf{v}_i\in \mathbb{R}^{N^2-1}$ \emph{satisfying Eq.\
(\ref{rpuro}) and weights $p_i$ satisfying $0 \leq p_i \leq 1$,
$\sum_i p_i = 1$ such that}
\begin{equation}\label{separable2}
T=p_i\textbf{u}_i\,\textbf{v}_i ^t,\quad
\textbf{r}=p_i\textbf{u}_i,\quad \textbf{s}=p_i\textbf{v}_i\,.
\end{equation}

This allows us to derive the two theorems below, which provide,
respectively, a necessary condition and a sufficient condition for
separability. We will make use of the Ky Fan norm $||\cdot||_{KF}$,
which is commonly used in Matrix Analysis (the reader who is not
familiarized with this issue can consult for example \cite{HorJoh}).
We recall that the singular value decomposition theorem ensures that
every matrix $A\in \mathbb{C}^{m\times n}$ admits a factorization of
the form $A=U\Sigma V^\dag$ such that $\Sigma=(\sigma_{ij})\in
\mathbb{R}_+^{m\times n}$ with $\sigma_{ij}=0$ whenever $i\neq j$,
and $U\in \mathbb{C}^{m\times m}$, $V\in \mathbb{C}^{n\times n}$ are
unitary matrices. The Ky Fan matrix norm is defined as the sum of
the singular values 
$\sigma_i\equiv\sigma_{ii}$,
\begin{equation}\label{kyfan}
||A||_{KF}=\sum_{i=1}^{\min\{m,n\}}\sigma_i=\textrm{Tr}\sqrt{A^\dag
A}.
\end{equation}
This norm has previously been used in the context of the
separability problem, though in a different way, in the CCNR
criterion.

\vspace*{12pt} \noindent {\bf Theorem~1:} If a bipartite state of
$M\times N$ dimensions with Bloch representation
(\ref{bipartitebloch}) is separable, then
\begin{equation}\label{teorema1}
||T||_{KF}\leq\sqrt{\frac{MN(M-1)(N-1)}{4}}
\end{equation}
must hold.

\vspace*{12pt} \noindent {\bf Proof:} Since $T$ has to admit a
decomposition of the form (\ref{separable2}) with
\begin{equation}
||\textbf{u}_i||_2=\sqrt{\frac{M(M-1)}{2}},\quad||\textbf{v}_i||_2=\sqrt{\frac{N(N-1)}{2}},
\end{equation}
we must have
\begin{equation}
||T||_{KF}\leq p_i||\textbf{u}_i\,\textbf{v}_i
^t||_{KF}=p_i\sqrt{\frac{MN(M-1)(N-1)}{4}}||\textbf{n}_i\,\tilde{\textbf{n}}_i
^t||_{KF},
\end{equation}
where $\textbf{n}_i,\tilde{\textbf{n}}_i$ are unit vectors. Thus,
$||\textbf{n}_i\,\tilde{\textbf{n}}_i ^t||_{KF}=1$ $\forall i$ and
the result follows. \hfill$\square$

\vspace*{12pt}

As said before, $T$ contains all the information about the
correlations, so that $||T||_{KF}$ measures in a certain sense the
size of these correlations. In this way, Theorem 1 has a clear
physical meaning: there is an upper bound to the correlations
contained in a separable state. $||T||_{KF}$ is a consistent measure
of the correlations since it is left invariant local changes of
basis, i.e. it is invariant under local unitary transformations of
the density operator. This fact was mentioned in \cite{Hor2} when
$M=N=2$; in the next proposition we give a general proof.

\vspace*{12pt} \noindent {\bf Proposition~2:} Let $U_A$ ($U_B$)
denote a unitary transformation acting on subsystem $A$ ($B$). If
\begin{equation}\label{uni}
\rho'=\big(U_A\otimes U_B\big)\rho\left(U_A^\dag\otimes
U_B^\dag\right),
\end{equation}
then $||T'||_{KF}=||T||_{KF}$.

\vspace*{12pt} \noindent {\bf Proof:} Let $\rho_A$ and $\rho'_A$
denote density operators acting on $H_A\simeq\mathbb{C}^M$ such that
$\rho'_A=U_A\rho_A U_A^\dag$. Since $||\cdot||_{HS}$ is unitarily
invariant we have that $||\rho_A||_{HS}=||\rho'_A||_{HS}$. But using
the orthogonality relation (\ref{ortogonalidad}) and Eq.\ (\ref{ri})
we find that
\begin{equation}
||\rho_A||_{HS}^2=\frac{1}{M}\left(1+\frac{2}{M}||\textbf{r}||_2^2\right),
\end{equation}
hence $||\textbf{r}||_2=||\textbf{r}'||_2$. This implies that the
coherence vectors of different realizations of the same state are
related by a rotation, i.e. there exists a rotation $O_A$ acting on
$\mathbb{R}^{M^2-1}$ such that $\textbf{r}'=O_A\textbf{r}$. This
means that
\begin{equation}
U_Ar_i\lambda_iU_A^\dag=\left(O_A\textbf{r}\right)_i\lambda_i.
\end{equation}
Now, when a bipartite state $\rho$ is subjected to a product unitary
transformation (\ref{uni}) there will be rotations $O_A$ acting on
$\mathbb{R}^{M^2-1}$ and $O_B$ acting on $\mathbb{R}^{N^2-1}$ such
that
\begin{equation}\label{trans}
\textbf{r}'=O_A\textbf{r},\quad\textbf{s}'=O_B\textbf{s},\quad
T'=O_ATO_B^\dag.
\end{equation}
Thus, the result follows taking into account that $||\cdot||_{KF}$
is unitarily invariant.
\hfill$\square$

\vspace*{12pt}

The characterization of the separability problem given in Eq.\
(\ref{separable2}) suggests the possibility of obtaining a
sufficient condition for separability using a constructive proof.
One such condition is stated in the following proposition.

\vspace*{12pt} \noindent {\bf Proposition~3:} If a bipartite state
of $M\times N$ dimensions with Bloch representation
(\ref{bipartitebloch}) satisfies
\begin{equation}\label{proposition3}
\sqrt{\frac{2(M-1)}{M}}||\textbf{r}||_2+\sqrt{\frac{2(N-1)}{N}}||\textbf{s}||_2+\sqrt{\frac{4(M-1)(N-1)}{MN}}||T||_{KF}\leq1,
\end{equation}
then it is a separable state.

\vspace*{12pt} \noindent {\bf Proof:} Let T have the singular value
decomposition $T=\sigma_i\textbf{u}_i\,\textbf{v}_i ^t$, with
$||\textbf{u}_i||_2=||\textbf{v}_i||_2=1$. If we define
\begin{equation}
\widetilde{\textbf{u}}_i=\sqrt{\frac{M}{2(M-1)}}\textbf{u}_i,\quad\widetilde{\textbf{v}}_i=\sqrt{\frac{N}{2(N-1)}}{\textbf{v}}_i,
\end{equation}
we can rewrite
\begin{equation}
T=\sqrt{\frac{4(M-1)(N-1)}{MN}}\sigma_i\widetilde{\textbf{u}}_i\,\widetilde{\textbf{v}}_i
^t.
\end{equation}
Then, if condition (\ref{proposition3}) holds, we can decompose
$\rho$ as the following convex combination of the density matrices
$\varrho_i$, $\varrho'_i$, $\rho_r$, $\rho_s$ and
$\frac{1}{MN}I_{MN}$,
\begin{eqnarray}\nonumber
\rho=
\sqrt{\frac{4(M-1)(N-1)}{MN}}\frac{1}{2}\sigma_i(\varrho_i+\varrho'_i)+\sqrt{\frac{2(M-1)}{M}}||\textbf{r}||_2\rho_r
+\sqrt{\frac{2(N-1)}{N}}||\textbf{s}||_2\rho_s\\\label{decompositionth2}
+\left(1-\sqrt{\frac{2(M-1)}{M}}||\textbf{r}||_2-\sqrt{\frac{2(N-1)}{N}}||\textbf{s}||_2-\sqrt{\frac{4(M-1)(N-1)}{MN}}||T||_{KF}\right)\frac{I_{MN}}{MN},
\end{eqnarray}
where $\varrho_i$, $\varrho'_i$, $\rho_r$ and $\rho_s$  are such
that
$$\textbf{r}_i=\widetilde{\textbf{u}}_i,\quad
\textbf{s}_i=\widetilde{\textbf{v}}_i,\quad
T_i=\widetilde{\textbf{u}}_i\,\widetilde{\textbf{v}}_i ^t,$$

$$\textbf{r}'_i=-\widetilde{\textbf{u}}_i,\quad
\textbf{s}'_i=-\widetilde{\textbf{v}}_i,\quad
T'_i=\widetilde{\textbf{u}}_i\,\widetilde{\textbf{v}}_i ^t,$$

$$\textbf{r}_r=\sqrt{\frac{M}{2(M-1)}}\frac{\textbf{r}}{||\textbf{r}||_2},\quad
\textbf{s}_r=0,\quad T_r=0,$$
$$\textbf{r}_s=0,\quad
\textbf{s}_s=\sqrt{\frac{N}{2(N-1)}}\frac{\textbf{s}}{||\textbf{s}||_2},\quad
T_s=0.$$ Notice that by virtue of Eq.\ (\ref{inclusion}) all the
above coherence vectors belong to the corresponding Bloch spaces
and, therefore, the reductions of $\varrho_i$, $\varrho'_i$,
$\rho_r$ and $\rho_s$ constitute density matrices. Moreover, all
these matrices satisfy condition (\ref{pureseparable}), hence they
are equal to the tensor product of their reductions. Therefore, they
constitute density matrices and they are separable, and so must be
$\rho$.\hfill$\square$

\vspace*{12pt}

One could ask whether Proposition 3 can be strengthened using a
condition more involved than Eq.\ (\ref{proposition3}). As we shall
see in the following theorem, the answer is positive.

\vspace*{12pt} \noindent {\bf Theorem~2:} Let
\begin{equation}\label{c}
c=\max\left\{\sqrt{\frac{2(M-1)}{M}}||\textbf{r}||_2,\sqrt{\frac{2(N-1)}{N}}||\textbf{s}||_2\right\}.
\end{equation}
If a bipartite state of $M\times N$ dimensions with Bloch
representation (\ref{bipartitebloch}) such that $c\neq0$ satisfies
\begin{equation}\label{teorema2}
c+\sqrt{\frac{4(M-1)(N-1)}{MN}}\left|\left|T-\frac{\textbf{r\,s}^t}{c}\right|\right|_{KF}\leq1,
\end{equation}
then it is a separable state.

\vspace*{12pt} \noindent {\bf Proof:} On the analogy of the proof of
Proposition 3, let $T-\frac{\textbf{r\,s}^t}{c}$ have the singular
value decomposition $\sigma'_i\textbf{x}_i\,\textbf{y}_i ^t$, where
$||\textbf{x}_i||_2=||\textbf{y}_i||_2=1$. If we define
\begin{equation}
\widetilde{\textbf{x}}_i=\sqrt{\frac{M}{2(M-1)}}\textbf{x}_i,\quad\widetilde{\textbf{y}}_i=\sqrt{\frac{N}{2(N-1)}}{\textbf{y}}_i,
\end{equation}
we can rewrite
\begin{equation}
T-\frac{\textbf{r\,s}^t}{c}=\sqrt{\frac{4(M-1)(N-1)}{MN}}\sigma'_i\widetilde{\textbf{x}}_i\,\widetilde{\textbf{y}}_i
^t.
\end{equation}
Now, if condition (\ref{teorema2}) holds we can decompose $\rho$ in
separable states as
\begin{eqnarray}\nonumber
\rho&=&
\sqrt{\frac{4(M-1)(N-1)}{MN}}\frac{1}{2}\sigma'_i(\varrho_i+\varrho'_i)
+c\rho_{rs}\\
 &+&
 \left(1-c-\sqrt{\frac{4(M-1)(N-1)}{MN}}\left|\left|T-\frac{\textbf{r\,s}^t}{c}\right|\right|_{KF}\right)\frac{1}{MN}I_{MN},
\end{eqnarray}
where $\varrho_i$, $\varrho'_i$ and $\rho_{rs}$ are such that
$$\textbf{r}_i=\widetilde{\textbf{x}}_i,\quad
\textbf{s}_i=\widetilde{\textbf{y}}_i,\quad
T_i=\widetilde{\textbf{x}}_i\,\widetilde{\textbf{y}}_i ^t,$$

$$\textbf{r}'_i=-\widetilde{\textbf{x}}_i,\quad
\textbf{s}'_i=-\widetilde{\textbf{y}}_i,\quad
T'_i=\widetilde{\textbf{x}}_i\,\widetilde{\textbf{y}}_i ^t,$$

$$\textbf{r}_{rs}=\frac{\textbf{r}}{c},\quad
\textbf{s}_{rs}=\frac{\textbf{s}}{c},\quad
T_{rs}=\frac{\textbf{r\,s}^t}{c^2}.$$ As in the previous proof, and
since
$$\frac{\textbf{r}}{c}\leq\sqrt{\frac{M}{2(M-1)}}\frac{\textbf{r}}{||\textbf{r}||_2},\quad\frac{\textbf{s}}{c}\leq\sqrt{\frac{N}{2(N-1)}}\frac{\textbf{s}}{||\textbf{s}||_2},$$
all these coherence vectors belong to the corresponding Bloch
spaces, and $\varrho_i$, $\varrho'_i$ and $\rho_{rs}$ satisfy
(\ref{pureseparable}).\hfill$\square$

\vspace*{12pt}

Notice that the use of the triangle inequality in Eq.\
(\ref{teorema2}) clearly shows that Theorem 2 is stronger than
Proposition 3. Nevertheless, Proposition 3 provides the right way to
understand the limit $c\rightarrow0$ in Theorem 2. The proof of
these two results is constructive, so for the states that fulfill
Eqs.\ (\ref{proposition3}) and/or (\ref{teorema2}) they provide a
decomposition in separable states. These states are in general not
pure, but they are equal to the tensor product of their reductions,
so to obtain a decomposition in product states as in Eq.\
(\ref{separable}) simply apply the spectral decomposition to the
reductions of $\varrho_i$, $\varrho'_i$, $\rho_r$, $\rho_s$ and/or
$\rho_{rs}$. 

\vspace*{12pt} \noindent {\bf Remark~2:} The conditions of
Proposition 3 and Theorem 2 depend only on $\textbf{r}$,
$\textbf{s}$ and $T$. However, there can also be obtained sufficient
conditions for separability which include more parameters. For
example, one can derive the following sufficient condition, which
also depends on the singular value decomposition of $T$,
\begin{equation}\label{remark2}
\left|\left|\sqrt{\frac{N}{2(N-1)}}\textbf{r}-\sigma_i\textbf{u}_i\right|\right|_2+\left|\left|\sqrt{\frac{M}{2(M-1)}}\textbf{s}-\sigma_i\textbf{v}_i\right|\right|_2+||T||_{KF}\leq\sqrt{\frac{MN}{4(M-1)(N-1)}},
\end{equation}
since in this case $\rho$ admits a decomposition in separable states
as in Eq.\ (\ref{decompositionth2}) but with $\varrho'_i=\varrho_i$,
$$\textbf{r}_r=\sqrt{\frac{M}{2(M-1)}}\frac{\textbf{r}-\sqrt{\frac{2(N-1)}{N}}\sigma_i\textbf{u}_i}{\left|\left|\textbf{r}-\sqrt{\frac{2(N-1)}{N}}\sigma_i\textbf{u}_i\right|\right|_2}\textrm{ and }
\textbf{s}_s=\sqrt{\frac{N}{2(N-1)}}\frac{\textbf{s}-\sqrt{\frac{2(M-1)}{M}}\sigma_i\textbf{v}_i}{\left|\left|\textbf{s}-\sqrt{\frac{2(M-1)}{M}}\sigma_i\textbf{v}_i\right|\right|_2}.$$
However, it seems reasonable to expect that condition
(\ref{remark2}) will be stronger than those of Proposition 3 and
Theorem 2 in few cases.

\vspace*{12pt}

For a restricted class of states the conditions of Theorem 1 and
Proposition 3 take the same form, thus providing a necessary and
sufficient condition which is equivalent to that of \cite{Hor2}:

\vspace*{12pt} \noindent {\bf Corollary~1:} A bipartite state of
qubits ($M=N=2$) with maximally mixed subsystems (i.e.
$\textbf{r}=\textbf{s}=0$) is separable if and only if
$||T||_{KF}\leq1$.

\vspace*{12pt}


\section{Efficacy of the New Criteria}

\subsection{Examples}
In what follows we provide examples of the usefulness of the
criteria derived in the previous section to detect entanglement. We
start by showing that Theorem 1 is strong enough to detect bound
entanglement.

\vspace*{12pt} \noindent {\it Example~1:} Consider the following
$3\times3$ PPT entangled state found in \cite{Ben}:
\begin{equation}\label{upb}
\rho=\frac{1}{4}\left(I_9-\sum_{i=0}^4|\psi_i\rangle\langle\psi_i|\right),
\end{equation}
where $|\psi_0\rangle=|0\rangle(|0\rangle-|1\rangle)/\sqrt{2}$,
$|\psi_1\rangle=(|0\rangle-|1\rangle)|2\rangle/\sqrt{2}$,
$|\psi_2\rangle=|2\rangle(|1\rangle-|2\rangle)/\sqrt{2}$,
$|\psi_3\rangle=(|1\rangle-|2\rangle)|0\rangle/\sqrt{2}$ and
$|\psi_4\rangle=(|0\rangle+|1\rangle+|2\rangle)(|0\rangle+|1\rangle+|2\rangle)/3$.
To construct the Bloch representation of this state we use as
generators of $SU(3)$ the Gell-Mann operators, which are a
reordering of those of Eqs.\
(\ref{generadoresa})-(\ref{generadoresc}),
\begin{equation}\label{gellmann}
\lambda_1=u_{01},\, \lambda_2=v_{01},\, \lambda_3=w_0,\,
\lambda_4=u_{02},\, \lambda_5=v_{02},\, \lambda_6=u_{12},\,
\lambda_7=v_{12},\, \lambda_8=w_1.
\end{equation}
Then, for the state (\ref{upb}) one readily finds
\begin{equation}
T=-\frac{1}{4}\left(
                \begin{array}{rrrrrrrr}
                  1 & 0 & 0 & 1 & 0 & 1 & 0 & \frac{\sqrt{27}}{2} \\
                  0 & 0 & 0 & 0 & 0 & 0 & 0 & 0 \\
                  -\frac{9}{4} & 0 & -\frac{9}{8} & 0 & 0 & 0 & 0 & \frac{\sqrt{27}}{8} \\
                  1 & 0 & 0 & 1 & 0 & 1 & 0 & 0 \\
                  0 & 0 & 0 & 0 & 0 & 0 & 0 & 0 \\
                  1 & 0 & -\frac{9}{4} & 1 & 0 & 1 & 0 & -\frac{\sqrt{27}}{4} \\
                  0 & 0 & 0 & 0 & 0 & 0 & 0 & 0 \\
                  -\frac{\sqrt{27}}{4} & 0 & \frac{\sqrt{27}}{8} & 0 & 0 & \frac{\sqrt{27}}{2} & 0 & -\frac{3}{8} \\
                \end{array}
              \right),
\end{equation}
so that $||T||_{KF}\simeq3.1603$, which violates condition
(\ref{teorema1}). Thus, using Theorem 1 we know that the state is
entangled.

\vspace*{12pt}

The above example proves that there exist cases in which Theorem 1
is stronger than the PPT criterion. One can see that this is not
true in general, not even for the $2\times2$ case.

\vspace*{12pt} \noindent {\it Example~2:} Consider the following
bipartite qubit state,
\begin{equation}\label{ejemplo2}
\rho_\pm=p|\psi^\pm\rangle\langle\psi^\pm|+(1-p)|00\rangle\langle00|\,,
\end{equation}
where $p \in [0,1]$ and
\begin{equation}
|\psi^\pm\rangle=\frac{1}{\sqrt{2}}\big(|01\rangle\pm|10\rangle\big).
\end{equation}
The Peres-Horodecki criterion establishes that state
(\ref{ejemplo2}) is separable iff $p=0$ \cite{Per}. For its Bloch
representation we use as generators of $SU(2)$ the standard Pauli
matrices $\sigma_x=u_{01}$, $\sigma_y=v_{01}$ and $\sigma_z=w_0$,
thus finding that
\begin{equation}
\rho_\pm=\frac{1}{4}(I_2\otimes I_2+(1-p)\sigma_z\otimes
I_2+(1-p)I_2\otimes\sigma_z\pm p\,\sigma_x\otimes\sigma_x\pm
p\,\sigma_y\otimes\sigma_y+(1-2p)\sigma_z\otimes\sigma_z).
\end{equation}
Therefore, $||T||_{KF}=2p+|1-2p|$, which implies that
$||T||_{KF}\leq1$ if $p\leq1/2$, so entanglement is detected only if
$p>1/2$.



\vspace*{12pt} \noindent {\it Example~3:} Werner states \cite{Wer}
in arbitrary dimensions ($M=N=D$) are those whose density matrices
are invariant under transformations of the form $\big(U\otimes
U\big)\rho\left(U^\dag\otimes U^\dag\right)$. They can be written as
\begin{equation}\label{wernerd}
\rho_W=\frac{1}{D^3-D}[(D-\phi)I_D\otimes I_D+(D\phi-1)V],
\end{equation}
where $-1\leq\phi\leq1$ and $V$ is the ``flip'' or ``swap'' operator
defined by
$V\varphi\otimes\widetilde{\varphi}=\widetilde{\varphi}\otimes\varphi$.
These states are separable iff $\phi\geq0$ \cite{Wer}. Using Eq.\
(\ref{T}) or inverting Eqs.\
(\ref{generadoresa})-(\ref{generadoresc}) we find that
\begin{equation}\label{V}
V=\sum_{i,j}|ij\rangle\langle ji|=\frac{1}{D}I_D\otimes
I_D+\frac{1}{2}\sum_lw_l\otimes
w_l+\frac{1}{2}\sum_{j<k}(u_{jk}\otimes u_{jk}+v_{jk}\otimes
v_{jk}),
\end{equation}
so that
\begin{equation}
\rho_W=\frac{1}{D^2}\left(I_D\otimes
I_D+\frac{D(D\phi-1)}{2(D^2-1)}\lambda_i\otimes\lambda_i\right),
\end{equation}
where $\lambda_i$ are the generators of $SU(D)$ defined as in Eqs.\
(\ref{generadoresa})-(\ref{generadoresc}). Then,
$||T||_{KF}=D|D\phi-1|/2$, so that Theorem 1 only recognizes
entanglement when $\phi\leq(2-D)/D$, while Proposition 3 guarantees
that the state is separable if $(D-2)/[D(D-1)]\leq\phi\leq1/(D-1)$.
When the latter condition holds, we can provide the decomposition in
product states. To illustrate the procedure, consider the Werner
state in, for simplicity, $2\times 2$ dimensions. In this case
$V=I_2\otimes I_2-2|\psi^-\rangle\langle\psi^-|$, and defining
$p=(1-2\phi)/3$ the state takes the simple form
\begin{equation}\label{werner}
\rho=\frac{1-p}{4}I_2\otimes
I_2+p|\psi^-\rangle\langle\psi^-|=\frac{1}{4}(I_2\otimes
I_2-p\,\sigma_x\otimes\sigma_x-p\,\sigma_y\otimes\sigma_y-p\,\sigma_z\otimes\sigma_z).
\end{equation}
From Corollary 1 we obtain that $\rho$ is separable iff $p\leq1/3$
as expected. From Proposition 3 we find that
\begin{equation}
\rho=\sum_{i=x,y,z}\sum_{j=1}^2\frac{p}{2}\rho_j^{(i)}+(1-3p)\frac{1}{4}(I_2\otimes
I_2),
\end{equation}
where
\begin{equation}
\rho_1^{(i)}=\frac{1}{4}(I_2\otimes I_2+\sigma_i\otimes
I_2-I_2\otimes\sigma_i-\sigma_i\otimes\sigma_i),\quad\rho_2^{(i)}=\frac{1}{4}(I_2\otimes
I_2-\sigma_i\otimes I_2+I_2\otimes\sigma_i-\sigma_i\otimes\sigma_i).
\end{equation}
In this case we can reduce the number of product states in the
decomposition to 8 by noticing that
$\rho_1^{(i)}=|01\rangle_{i}\langle01|$ and
$\rho_2^{(i)}=|10\rangle_i\langle10|$, where
$\{|0\rangle_i,|1\rangle_i\}$ denote the eigenvectors of $\sigma_i$,
so that, for instance,
\begin{align}
\rho&=\sum_{i=x,y}\frac{p}{2}(|01\rangle_{i}\langle01|+|10\rangle_i\langle10|)+\frac{1-p}{4}(|01\rangle_z\langle01|+|10\rangle_z\langle10|)+\nonumber\\
&+\frac{1-3p}{4}(|00\rangle_z\langle00|+|11\rangle_z\langle11|).
\end{align}
It is known, however, that a separable bipartite qubit state admits
a decomposition in a number of product states less than or equal to
4 \cite{Woo,San}.

\vspace*{12pt}

\noindent {\it Example~4:} Isotropic states \cite{iso} in arbitrary
dimensions ($M=N=D$) are invariant under transformations of the form
$\big(U\otimes U^\ast\big)\rho\left(U^\dag\otimes
U^{\ast\dag}\right)$. They can be written as mixtures of the
maximally mixed state and the maximally entangled state
\begin{equation}\label{maxentangled}
|\Psi\rangle=\frac{1}{\sqrt{D}}\sum_{a=0}^{D-1}|aa\rangle,
\end{equation}
so they read\footnote{In the two-qubit case the Werner ($U\otimes U$
invariant) states (\ref{werner}) and isotropic ($U\otimes U^\ast$
invariant) states (\ref{isotropic}) are identical up to a local
unitary transformation. For this reason some authors refer to the
isotropic states as generalized Werner states, which might lead to
confusion.}
\begin{equation}\label{isotropic}
\rho=\frac{1-p}{D^2}I_D\otimes I_D+p|\Psi\rangle\langle\Psi|.
\end{equation}
These states are known to be separable iff $p\leq(D+1)^{-1}$
\cite{iso} (see also \cite{Run,Pit}). Their Bloch representation can
be easily found as in the Werner case, and it is given by
\begin{equation}
\rho=\frac{1}{D^2}\left(I_D\otimes
I_D+\frac{pD}{2}\sum_{i=1}^{(D+2)(D-1)/2}\lambda_i\otimes\lambda_i-\frac{pD}{2}\sum_{i=D(D+1)/2}^{D^2-1}\lambda_i\otimes\lambda_i\right),
\end{equation}
where, as before, $\lambda_i$ are the generators of $SU(D)$ defined
in Eqs.\ (\ref{generadoresa})-(\ref{generadoresc}). Now,
$||T||_{KF}=pD(D^2-1)/2$. Thus, Theorem 1 is strong enough to detect
all the entangled states ($||T||_{KF}\leq D(D-1)/2\Leftrightarrow
p\leq(D+1)^{-1}$), while Proposition 3 ensures that the states are
separable when $p\leq(D+1)^{-1}(D-1)^{-2}$.


\subsection{Comparison with the CCNR criterion}

Let $\rho$ be written in terms of the canonical basis
$\{E_{ij}\otimes E_{kl}\}$ of $HS(H_A\otimes H_B)$ as
\begin{equation}
\rho=c_{ijkl}E_{ij}\otimes E_{kl}.
\end{equation}
The computable cross-norm criterion, proposed by O. Rudolph (see
\cite{Rud1,Rud2} and references therein), states that for all
separable states the operator $U(\rho)$ acting on $HS(H_A\otimes
H_B)$ defined by
\begin{equation}
U(\rho)\equiv c_{ijkl}|E_{ij}\rangle\langle E_{kl}|,
\end{equation}
where $|E_{mn}\rangle$ denotes the ket vector with respect to the
inner product in $HS(H_A)$ or $HS(H_B)$, is such that
$||U(\rho)||_{KF}\leq1$. Soon after, K. Chen and L.-A. Wu derived
the realignment method \cite{Che}, which yields the same results as
the cross-norm criterion from simple matrix analysis. Basically, it
states that a certain realigned version of a separable density
matrix cannot have Ky Fan norm greater than one, thus providing a
simple way to compute this condition. This is why we refer to it as
the CCNR criterion. Like Theorem 1, it is able to detect all
entangled isotropic states and recognizes entanglement for the same
range of Werner states \cite{Rud1}. Although being weaker than the
PPT criterion in $2\times2$ dimensions, it is also capable of
detecting bound entangled states. However, the CCNR criterion
detects optimally the entanglement of the state of Example 2
\cite{Rud1}, so one could think that it is stronger than Theorem 1.
To check this possibility and to evaluate the ability of bound
entanglement detection of Theorem 1, we have programmed a routine
that generates $10^6$ random $3\times3$ PPT entangled states
following \cite{Bru}. Our theorem detected entanglement in about
$4\%$ of the states while the CCNR criterion recognized $18\%$ of
the states as entangled. Moreover, every state detected by Theorem 1
was also detected by the CCNR criterion. This suggests that the CCNR
criterion is stronger than Theorem 1 when $M=N$. We will show that
this is indeed the case, but we will also see that this is not true
when $M\neq N$. First we will prove the following lemma:

\vspace*{12pt} \noindent {\bf Lemma~1:}
$$\left|\left|\left(
                                                        \begin{array}{cc}
                                                          A & B \\
                                                          C & D \\
                                                        \end{array}
                                                      \right)\right|\right|_{KF}\geq
||A||_{KF}+||D||_{KF},$$ where $A,B,C,D$ are complex matrices of
adequate dimensions.

\vspace*{12pt} \noindent {\bf Proof:} Let $A$ and $D$ have the
singular value decompositions $A=U_A\Sigma_AV_A^{\dag}$ and
$D=U_D\Sigma_DV_D^{\dag}$. It is clear from the definition that the
Ky Fan norm is unitarily invariant. Therefore, we have that
\begin{align}
\left|\left|\left(
                                                        \begin{array}{cc}
                                                          A & B \\
                                                          C & D \\
                                                        \end{array}
                                                      \right)\right|\right|_{KF}&=\left|\left|\left(
                                                                                               \begin{array}{cc}
                                                                                                 U_A^{\dag} & 0 \\
                                                                                                 0 & U_D^{\dag} \\
                                                                                               \end{array}
                                                                                             \right)
\left(
                                                        \begin{array}{cc}
                                                          A & B \\
                                                          C & D \\
                                                        \end{array}
                                                      \right)\left(
                                                               \begin{array}{cc}
                                                                 V_A & 0 \\
                                                                 0 & V_D \\
                                                               \end{array}
                                                             \right)
\right|\right|_{KF}\nonumber\\
& \geq \textrm{Tr }\Sigma_A + \textrm{Tr }\Sigma_D,
\end{align}
where we have used that $||X||_{KF}\geq \textrm{Tr } X$, which is a
direct consequence of the following characterization of the Ky Fan
norm (see Eq.\ (3.4.7) in \cite{HorJoh}):
\begin{equation}
||X||_{KF}=\max\{|\textrm{Tr } XU|: U \textrm{ is unitary}\}.
\end{equation}
\hfill$\square$

\vspace*{12pt} \noindent {\bf Proposition~4:} In the case of states
with maximally mixed subsystems Theorem 1 is stronger than the CCNR
criterion when $M\neq N$, while when $M=N$ they are equivalent.

\vspace*{12pt} \noindent {\bf Proof:} When $\textbf{r}=\textbf{s}=0$
we have that
\begin{equation}
U(\rho)=\frac{1}{MN}(|I_M\rangle\langle
I_N|+t_{ij}|\lambda_i\rangle\langle\tilde{\lambda}_j|).
\end{equation}
Since the matrix associated to the operator $U(\rho)$ is in this
case block-diagonal we find that
\begin{align}
||U(\rho)||_{KF}&=\frac{1}{\sqrt{MN}}\left|\left|\frac{|I_M\rangle}{\sqrt{M}}\frac{\langle
I_N|}{\sqrt{N}}\right|\right|_{KF}+\frac{2}{MN}\left|\left|t_{ij}\frac{|\lambda_i\rangle}{\sqrt{2}}
\frac{\langle\tilde{\lambda}_j|}{\sqrt{2}}\right|\right|_{KF}\nonumber\\
& =\frac{1}{\sqrt{MN}}+\frac{2}{MN}||T||_{KF}.
\end{align}
Thus, for states with maximally mixed subsystems the CCNR criterion
is equivalent to
\begin{equation}
||T||_{KF}\leq\frac{\sqrt{MN}(\sqrt{MN}-1)}{2},
\end{equation}
from which the statement readily follows. \hfill$\square$

\vspace*{12pt} \noindent {\bf Proposition~5:} The CCNR criterion is
stronger than Theorem 1 when $M=N$.

\vspace*{12pt} \noindent {\bf Proof:} Since in this case in general
$\textbf{r},\textbf{s}\neq0$, the matrix associated to the operator
$U(\rho)$ is no longer block-diagonal. Hence, using Lemma 1, we now
have that
\begin{equation}
||U(\rho)||_{KF}\geq\frac{1}{N}+\frac{2}{N^2}||T||_{KF},
\end{equation}
which proves the result considering that in the $M=N$ case the
condition of Theorem 1 can be written as
\begin{equation}
\frac{1}{N}+\frac{2}{N^2}||T||_{KF}\leq1.
\end{equation}
\hfill$\square$

\vspace*{12pt}

Proposition 4 explains why both criteria yield the same results for
Werner and isotropic states. However, since $T$ is diagonal in these
cases, the computations are much simpler in our formalism than in
that of the CCNR criterion. Furthermore, when $M\neq N$ we have
explicitly constructed entangled states which are detected by
Theorem 1 but not by the CCNR criterion. Regrettably, Theorem 1 is
not able to detect the PPT entangled states in $2\times4$ dimensions
constructed by P. Horodecki in \cite{Hor97}.

\section{Summary and Conclusions}

We have used the Bloch representation of density matrices of
bipartite quantum systems in arbitrary dimensions $M\times N$, which
relies on two coherence vectors $\textbf{r}\in \mathbb{R}^{M^2-1}$,
$\textbf{s}\in \mathbb{R}^{N^2-1}$ and a correlation matrix $T\in
\mathbb{R}^{(M^2-1)\times (N^2-1)}$, to study their separability.
This approach has led to an alternative formulation of the
separability problem, which has allowed us to characterize entangled
pure states (Proposition 1), and to derive a necessary condition
(Theorem 1) and three sufficient conditions (Proposition 3, Theorem
2 and Remark 2) for the separability of general states. In the case
of bipartite systems of qubits with maximally mixed subsystems
Theorem 1 and Proposition 3 take the same form, thus yielding a
necessary and sufficient condition for separability. We have shown
that, despite being weaker than the PPT criterion in $2\times2$
dimensions, Theorem 1 is strong enough to detect PPT entangled
states. We have also shown that it is capable of recognizing all
entangled isotropic states in arbitrary dimensions but not all
Werner states, like the CCNR criterion. Although the CCNR criterion
turns out to be stronger than Theorem 1 when $M=N$, we have also
proved that our theorem is stronger than the CCNR criterion for
states with maximally disordered subsystems when $M\neq N$.
Therefore, although Theorem 1 does not fully characterize
separability, we believe that in combination with the above criteria
it can improve our ability to understand and detect entanglement.
Theorem 2, together with Proposition 3 (which is weaker save for the
limiting case $c=0$) and the result of Remark 2 (which is more
involved), offers a sufficiency test of separability, which, as a
by-product, provides a decomposition in product states of the states
that satisfy its hypothesis. $||T||_{KF}$ acts as a measure of the
correlations inside a bipartite state and it is left invariant under
local unitary transformations of the density matrix. This suggests
the possibility of considering it as a rough measure of
entanglement, as in the case of the realignment method \cite{Che}.
We think that this subject deserves further study. We also believe
that a deeper understanding of the geometrical character of the
Bloch-vector space could lead to an improvement of the separability
conditions presented here.

\section*{Acknowledgements} \noindent The author is very much
indebted to Jorge S\'anchez-Ruiz for useful comments and,
particularly, for his suggestion of the present form of Theorem 2,
which improved on a previous version of the theorem. He is also very
thankful to Otfried G\"{u}hne for discussions and remarks that led
to substantial improvements in Section 4.2. Financial support by
Universidad Carlos III de Madrid and Comunidad Aut\'onoma de Madrid
(project No. UC3M-MTM-05-033) is gratefully acknowledged.


\end{document}